\documentstyle[12pt,equation]{article}
\begin{document}

\newcommand{\tanb}{\mbox{$\tan \! \beta$}}
\newcommand{\msb}{\mbox{$m_{\tilde{b}_L}$}}
\newcommand{\mst}{\mbox{$m_{\tilde{t}_1}$}}
\newcommand{\mstt}{\mbox{$m_{\tilde{t}_2}$}}
\newcommand{\mstlr}{\mbox{$m_{\tilde{t}_{L,R}}$}}
\newcommand{\mstsq}{\mbox{$m^2_{\tilde{t}_1}$}}
\newcommand{\stst}{\mbox{$(\tilde{t}_1 \tilde{t}_1^*)$}}
\newcommand{\sigst}{\mbox{$\sigma_{\tilde{t}_1}$}}
\newcommand{\gamgam}{\mbox{$\gamma \gamma$}}
\newcommand{\st}{\mbox{$\tilde{t}_1$}}
\newcommand{\stt}{\mbox{$\tilde{t}_2$}}
\newcommand{\stl}{\mbox{$\tilde{t}_L$}}
\newcommand{\str}{\mbox{$\tilde{t}_R$}}
\newcommand{\ww}{\mbox{$W^+W^-$}}
\newcommand{\mat}{\mbox{${\cal M}^2_{\tilde{t}}$}}
\newcommand{\be}{\begin{equation}}
\newcommand{\ee}{\end{equation}}
\newcommand{\een}{\end{subequations}}
\newcommand{\ben}{\begin{subequations}}
\newcommand{\beq}{\begin{eqalignno}}
\newcommand{\eeq}{\end{eqalignno}}
\renewcommand{\thefootnote}{\fnsymbol{footnote} }
\noindent
\begin{flushright}
MAD/PH/782\\
KEK--TH--368\\
KEK Preprint 93--106\\
August 1993
\end{flushright}
\vspace{1.5cm}
\pagestyle{empty}
\begin{center}
{\Large \bf Scalar Stoponium}\footnote{Contribution to the Proceedings of the
Madison--Argonne Hadron Collider Workshop, March to June 1993} \\
\vspace{5mm}
Manuel Drees\footnote{Heisenberg Fellow}\\
{\em Physics Department, University of Wisconsin, Madison, WI 53706, USA}\\
Mihoko M. Nojiri\footnote{E--mail: NOJIRIN@JPNKEKVX}\\
{\em Theory Group, KEK, Oho 1--1, Tsukuba, Ibaraki 305, Japan}
\end{center}

\begin{abstract}
\noindent
We study the decays of a scalar \stst\ bound state \sigst, where \st\ is the
lighter stop eigenstate. If \st\ has no tree--level 2--body decays, the
dominant decay modes of \sigst\ are $gg$ or, if $m_h < \mst \ll \mstt$, a pair
of light scalar Higgs bosons $h$. The best signal for \sigst\ production at
hadron colliders is probably its decay into two photons.
\end{abstract}
\vspace*{1cm}
\clearpage
\noindent
\setcounter{footnote}{0}
\pagestyle{plain}
\setcounter{page}{1}
It is by now quite well known that the lighter scalar top (stop) eigenstate
\st\ is expected to be lighter than the superpartners of the first two
generations of quarks, and might even be the lightest colored sparticle
\cite{1,2}. There are two reasons for this: Since the top quark is heavy,
$m_t \geq 110$ GeV, mixing between the superpartners \stl, \str\ of left-- and
right--handed top quarks cannot be neglected, in contrast to the superpartners
of light quarks. Furthermore, if we assume all squarks to have the same mass
at some very high (GUT, string or Planck) scale, radiative corrections
\cite{3} will reduce the masses of \stl\ and \str\ relative to those of the
other squarks.

Open \st\ pair production at $e^+e^-$ \cite{4} and $\bar{p} p$ \cite{5}
colliders has been discussed previously. Here we study possible signals for
the production of a scalar \stst\ bound state \sigst\ within the Minimal
Supersymmetric Standard Model (MSSM) \cite{7}. In ref.\cite{6} it has been
pointed out that the branching ratio for $\sigst \rightarrow hh$ might be
large, where $h$ is the lighter scalar Higgs boson. Very recently it has been
claimed \cite{8} that $\sigst \rightarrow \ww$ can have a very large branching
ratio, which might give rise to interesting signals at hadron supercolliders.
This motivated us to compute all potentially large branching ratios of \sigst.
We basically agree with the results of ref.\cite{6}, but were unable to
reproduce those of ref.\cite{8}. The cleanest signal for \sigst\ production
at hadron colliders arises from its 2--photon decay, giving rise to a peak
in the \gamgam\ invariant mass spectrum.

The starting point of our discussion is the stop mass matrix. Following the
convention of ref.\cite{9} we write it as (in the basis \stl, \str):
\be \label{e1}
\mat = \mbox{$ \left( \begin{array}{cc}
m^2_{\tilde{t}_L} + m_t^2 + 0.35 D_Z & - m_t (A_t + \mu \cot \! \beta) \\
- m_t (A_t + \mu \cot \! \beta ) & m^2_{\tilde{t}_R} + m_t^2 + 0.16 D_Z
\end{array} \right) $}. \ee
Here, $D_Z = M_Z^2 \cos\!2 \beta$ with $\tanb = \langle \bar{H^0} \rangle /
\langle H^0 \rangle$ as usual \cite{7}, $\mu$ is the supersymmetric Higgs(ino)
mass parameter, $A_t$ a trilinear soft supersymmetry breaking parameter, and
$m^2_{\tilde{t}_{L,R}}$ the nonsupersymmetric contributions to the squared
masses of the \stl, \str\ current states. The diagonalization of \mat\ is
straightforward. We find for the lighter eigenstate $\st \equiv \cos \!
\theta_t \stl + \sin \! \theta_t \str$: \ben \label{e2} \beq
\mstsq &= \frac{1}{2} \left[ m^2_{LL} + m^2_{RR} - \sqrt{ \left( m^2_{LL}
 - m^2_{RR} \right)^2 + 4 m^4_{LR} } \right]; \label {e2a} \\
\tan \! \theta_t &= \frac {\mstsq - m^2_{LL}} {m^2_{LR}}, \label{e2b}
\eeq \een
where $m^2_{LL,LR,RR}$ refers to the $LL, \ RR$ and $LR$ elements of \mat.
While the gauge interactions of \st\ only depend on $\theta_t$, the couplings
of stop squarks to Higgs bosons depend on all parameters entering
eq.(\ref{e1}); all these quantities therefore have to be specified before
\sigst\ branching ratios can be computed.

There are two different kinds of \sigst\ decays: Single stop decays and
annihilation decays. In the first case either \st\ or $\tilde{t}_1^*$
decays, leaving the other squark behind. We assume that the gluino is too
heavy to be produced in these decays. In general we then have to consider
the following decay modes: \ben \label{e3} \beq
\st &\rightarrow b \tilde{W}_i, i=1,2 ; \label{e3a} \\
\st &\rightarrow t \tilde{Z}_i, i=1,\dots,4 ; \label {e3b} \\
\st &\rightarrow c \tilde{Z}_i, i=1,\dots,4 , \label {e3c}
\eeq \een
where $\tilde{W}_i$ ($\tilde{Z}_i$) denotes a generic chargino (neutralino)
state. The decays (\ref{e3a},\ref{e3b}) occur at tree level and with full
gauge or top Yukawa strength. It has been shown in ref.\cite{2} that
(\ref{e3c}) is the dominant \st\ decay mode if the first two modes are
kinematically forbidden. However, this last decay only occurs at 1--loop level
and necessitates flavor mixing; it is therefore suppressed relative to the
tree--level decays by a factor $|\epsilon|^2 \simeq 10^{-7}$ \cite{2}. We will
see below that the mode (\ref{e3c}) can therefore be neglected in the
discussion of \sigst\ decays. The widths of (\ref{e3a},\ref{e3b}) can be
computed using the couplings of refs.\cite{7}; the decay width of \sigst\ is
twice that of \st.

In annihilation decays \st\ and $\tilde{t}_1^*$ annihilate into a
flavor and color singlet final state; this kind of decay is by far dominant
for the familiar $(c \bar{c})$ and $(b \bar{b})$ bound states (quarkonia). We
computed the widths for the following modes:
\ben \label{e4} \beq
\sigst &\rightarrow gg \label{e4a}; \\
\sigst &\rightarrow \ww; \label{e4b} \\
\sigst &\rightarrow ZZ; \label{e4c} \\
\sigst &\rightarrow Z \gamma; \label{e4d} \\
\sigst &\rightarrow \gamgam; \label{e4e} \\
\sigst &\rightarrow hh; \label{e4f} \\
\sigst &\rightarrow b \bar{b}; \label{e4g} \\
\sigst &\rightarrow t \bar{t}; \label{e4h} \\
\sigst &\rightarrow \tilde{Z}_i \tilde{Z}_j, i,j=1,\dots,4. \label{e4i}
\eeq \een
We computed the corresponding branching ratios using ``Method 2'' described in
the Appendix of ref.\cite{10}. Since \sigst\ is a scalar ($s-$wave) state, we
only need the \st\ velocity $v \rightarrow 0$ limit of the $\st \tilde{t}_1^*$
annihilation amplitudes leading to the final states of eqs.(\ref{e4}). In this
limit, reactions (\ref{e4a},\ref{e4d},\ref{e4e}) proceed via $t-$channel \st\
exchange as well as via 4--point ``butterfly'' interactions; (\ref{e4b})
proceeds via $\tilde{b}_L$ exchange\footnote{We ignore mixing in the
$\tilde{b}$ sector.}, a 4--point interaction as well as scalar Higgs exchange
in the $s-$channel, while (\ref{e4c},\ref{e4f}) involve \st\ or \stt\ exchange
in the $t-$channel, a 4--point coupling and Higgs exchange.\footnote{\stt\
exchange has not been included in ref.\cite{6}, where the reaction (\ref{e4f})
has been studied; this contribution is small compared to the \st\ exchange
term for parameters leading to a sizable $hh$ branching ratio.}
Processes (\ref{e4g},\ref{e4h}) involve $s-$channel scalar Higgs exchange and
$t-$channel chargino or neutralino exchange; note that the corresponding
matrix elements are proportional to the final state quark masses, so that
the width for (\ref{e4g}) is very small unless $\tanb \gg 1$. Finally,
(\ref{e4i}) proceeds via $s-$channel Higgs exchange or $t-$channel top
exchange.

In order to compute the decay widths for the processes (\ref{e4}) we have to
know the ``wave function at the origin'' $|\psi(0)|^2$, see ref.\cite{10}.
Recently the \mst\ dependence of this quantity has been parametrized in
ref.\cite{11}, for 4 different values of the QCD scale parameter $\Lambda$,
using a potential that reproduces the known quarkonium spectrum well; we use
their fit for $\Lambda = 0.2$ GeV.

In fig.1 we show examples for the branching ratios of processes (\ref{e4})
for relatively small $m_{\tilde{t}_{L,R}}$ (200 GeV). In addition to the
parameters appearing in eq.(\ref{e1}) we have to specify the SU(2) gaugino
mass $M_2$ (we assume $M_1 = 5/3 \tan^2 \theta_W M_2$ as usual) and the mass
$m_P$ of the pseudoscalar Higgs boson. This then determines all relevant
masses and couplings. We have included leading radiative corrections to the
scalar Higgs sector involving top--stop loops \cite{rad}.

In this figure we have assumed $M_2 = 100$ GeV leading to a mass of about
110 GeV for the lighter chargino. For $\mst > 115$ GeV the single stop decay
(\ref{e3a}) (not shown) opens up and quickly dominates the total decay width.
Indeed, in this region the total width of \sigst\ is comparable to its binding
energy. Our calculations are no longer reliable in this case, since we
assume that formation and decay of \sigst\ occur at very different time scales
so that they can be treated independently; one has to use methods developed
previously \cite{12} for $(t \bar{t})$ bound states instead. However, we can
conclude from fig. 1 that if the single stop decays (\ref{e3a},\ref{e3b}) are
allowed the branching ratios for final states that might be detectable at
hadron colliders (see below) are very small, less than $10^{-4}$.

In fig. 2 we have therefore varied $M_2$ along with \mst, so that the decays
(\ref{e3a},\ref{e3b}) remain closed for $\mst \leq |\mu|$. We have also
chosen larger values for $m_{\tilde{t}_{L,R}}$, with $m_{\tilde{t}_L} >
m_{\tilde{t}_R}$ as predicted by minimal supergravity models \cite{9}.
We see that now the $Br(\sigst \rightarrow hh)$ shoots up very rapidly once
this decay becomes kinematically allowed. The reason is that the $h \st \st$
coupling \cite{7} contains a term which, in the limit $m_P^2 \gg M_Z^2$,
is exactly proportional to the $LR$ element of \mat. Obviously this element
has to be large if \mst\ is to be much smaller than $m_{\tilde{t}_{L,R}}$. As
a result, the amplitude for $\st \tilde{t}_1^* \rightarrow hh$ is
proportional to $\left( m^2_{\tilde{t}_{L,R}}/M_W \mst \right)^2$ if
$m^2_{\tilde{t}_{L,R}} \gg \mstsq \gg m^2_h/2$. This explains the decrease
of the $hh$ branching ratio with increasing \mst, in spite of the increasing
phase space.

The branching ratios for \ww\ and $ZZ$ also increase quickly just beyond
threshold. However, they do not reach the level of the $hh$ branching ratio;
their amplitudes are at best $\propto \left( m_{\tilde{t}_L}/M_W \right)^2$,
if $m^2_{\tilde{t}_{L,R}} \gg \mstsq \gg M^2_W$. This can be understood from
the equivalence theorem \cite{13}, which states that amplitudes involving
longitudinal gauge bosons are equal to corresponding ones involving
pseudoscalar Goldstone bosons $G$, if the energy of the process is $\gg m_W$.
There is no diagonal $G \st \st$ coupling; a $G \st \stt$ coupling with
strength similar to the $h \st \st$ coupling does exist, but it only affects
\sigst\ decays via diagrams involving a {\em heavy} stop propagator. The
amplitude for $\sigst \rightarrow Z_L Z_L$ is therefore suppressed by a factor
$\left( \mst / \mstt \right)^2$ compared to the $hh$ amplitude. Similar
arguments apply for the $W^+_L W^-_L$ amplitude. Transverse $W$ and $Z$ bosons
are at best produced with ordinary (weak) gauge strength, and their couplings
can even be suppressed by $\tilde{t}$ mixing. Unlike ref.\cite{8} we therefore
never find the width into $WW$ to exceed the one into gluons. However, the
authors of ref.\cite{8} neglected $\tilde{t}$ mixing, and assumed that
$m_{\tilde{t}_L}$ can be varied independently of the mass of the left--handed
sbottom $\tilde{b}_L$. Since $\tilde{t}_L$ and $\tilde{b}_L$ reside in the
same $SU(2)$ doublet, this introduces a new source of explicit gauge symmetry
breaking, which renders the theory nonrenormalizable.

The curves of fig. 2 show a lot of structure around \mst=250 GeV. This is
because for the given choice of parameters the mass of the heavy scalar
Higgs $H$ is just above 500 GeV; the $s-$channel $H$ exchange diagrams
therefore become resonant, greatly enhancing the matrix elements for
$t \bar{t}, \ b \bar{b}, \ hh$ and $\tilde{Z}_i \tilde{Z}_j$. The enhancement
of the \ww\ and $ZZ$ final states is much weaker, since the $HVV$ couplings
($V=W,Z$) are small for $m_H^2 \gg M_Z^2$. If 2\mst\ is very close to
$m_H$ our treatment again breaks down; in this case the \stst\ bound states
mix with $H$ \cite{14}.

Finally, the curves of figs. 1,2 exhibit numerous minima resulting from
destructive interference between different contributions to the matrix
elements. In particular, for the \ww, $ZZ$ and $hh$ final states the
$t-$channel and 4--point coupling diagrams always contribute with opposite
signs. Since the size of the $t-$channel contributions increases with
increasing ratio $m_{\tilde{t}_{L,R}}/\mst$, the interference minima move
further away from threshold as \mstlr\ are increased. The $hh$ branching ratio
in fig. 2 has a second minimum due to interference with the $s-$channel $H$
exchange contribution. The $t \bar{t}, \ \tilde{Z}_1 \tilde{Z}_1$ and
$\tilde{Z}_1 \tilde{Z}_2$ final states also show destructive interference
between $s-$ and $t-$channel diagrams. It is important to note that (at least
in the limit $v \rightarrow 0$) usually only a single partial wave is
accessible in \sigst\ decays; if in addition only a single combination of
final state helicities can be produced, destructive interference can lead to a
vanishing total amplitude even far above threshold.

Clearly one could in principle learn a lot about the MSSM parameters by
studying \sigst\ branching ratios. In practice, however, even discovery
of \sigst\ may not be trivial. We focus here on $pp$ supercolliders.
The total cross section for \sigst\ production at the SSC is shown by the
solid line in fig. 3. Hadronic final states ($gg, \ b \bar{b}, t \bar{t}$)
will be useless for the discovery of \sigst\ at such colliders, due to the
enormous backgrounds. In ref.\cite{8} the use of the \ww\ final state was
advocated. However, we have seen above that $SU(2)$ gauge invariance implies
a rather small rate for this final state; besides, it is not clear to us how
the \ww\ invariant mass will be measured, since both $W$ bosons will have to
decay leptonically in order to suppress QCD backgrounds. The $ZZ$ final state
is very clean if both $Z$ bosons decay leptonically, but then the rate will
be very small even at the SSC ($<$ 5 events/year).

In ref.\cite{6} the use of $\sigst \rightarrow hh \rightarrow \tau^+ \tau^+
\tau^- \tau^-$ has been proposed. Since $Br(h \rightarrow \tau^+ \tau^-)
\simeq 8\%$ the 4 $\tau$ final state is also relatively rare ($Br < 6.4
\cdot 10^{-3}$). The real problem is that it is not clear how this final
state is to be identified experimentally in a hadronic environment. A pair
of isolated like--sign leptons would be a rather clean tag; assuming a $30\%$
detection efficiency per lepton one could get more than 50 events/year for
$\mst \leq 100$ GeV if the $hh$ final state dominates. The problem is that
the presence of at least 4 neutrinos in the final state makes it impossible to
reconstruct the \sigst\ mass. Given that there are sizable backgrounds
(e.g., $\sigma(pp \rightarrow ZZX \rightarrow  \tau^+ \tau^+
\tau^- \tau^- X) \simeq 3 \cdot 10^{-2}$ pb, leading to $\sim 300$
events/year) the identification of the 4 $\tau$ signal might be problematic.

Probably the most promising signal results from the decay $\sigst \rightarrow
\gamgam$. Searches for intermediate mass Higgs bosons will presumably require
good electromagnetic energy resolution for SSC detectors, so the
reconstruction of $m(\sigst)$ should be rather straightforward. The signal
would then be a bump in the \gamgam\ spectrum on top of the smooth
background. The dashed curve in fig. 3 shows the signal cross section for the
same parameters as in fig. 2. In ref.\cite{15} the minimal detectable $H
\rightarrow \gamgam$ cross section has been estimated. The result after 1
year of running can roughly be parametrized as $\sigma_1^{\rm min} =
0.05 {\rm pb} \cdot \left( m_H/100 \ {\rm GeV} \right)^{-1.36}$, and is shown
by
the dotted curve in fig. 3. The same result obviously also applies for
\sigst\ searches, since the final state is identical. Except for the vicinity
of the Higgs pole, where our calculation is not reliable anyway, our signal
cross section is always above this minimum. Even though the assumptions of
ref.\cite{15} might be somewhat optimistic, it is clear that for most
combinations of parameters \sigst\ should be readily observable via its
\gamgam\ decay already after one year of SSC running.

In summary, we have computed all potentially large branching ratios of a
scalar stoponium bound state \sigst. The dominant decay modes are $gg, \ hh$
or, if $m(\sigst) \simeq m_H, \ t \bar{t}$. The process $pp \rightarrow
\sigst X \rightarrow \gamgam X$ should be readily observable at the SSC,
{\em provided} that \st\ has no unsuppressed tree--level 2--body decays.

\subsection*{Acknowledgements}
We thank X. Tata for useful discussions and suggestions, H. Inazawa and T.
Morii for communications on ref.\cite{8}, and V. Barger for bringing
ref.\cite{6} to our attention. The work of M.D. was supported in part by the
U.S. Department of Energy under contract No. DE-AC02-76ER00881, and in part by
the Wisconsin Research Committee with funds granted by the Wisconsin Alumni
Research Foundation, as well as by a grant from the Deutsche
Forschungsgemeinschaft under the Heisenberg program.

\renewcommand{\theequation}{A.\arabic{equation}}
\setcounter{equation}{0}

\section*{Figure Captions}

\renewcommand{\labelenumi}{Fig.\arabic{enumi}}
\begin{enumerate}

\item 
Branching ratios for annihilation decays of \sigst\ listed in eqs.(\ref{e4}).
The range of \mst\ values shown results from varying $A_t$ between -310 and
-70 GeV. For $\mst > 115$ GeV single stop decays open up; in this region
annihilation decays into charginos and heavier neutralinos are also possible,
but they remain small, of order of the $\tilde{Z}_1 \tilde{Z}_{1,2}$ modes
shown in the figure.

\vspace*{5mm}
\item   
Branching ratios for annihilation decays of \sigst\ listed in eqs.(\ref{e4}).
The range of \mst\ values shown results from varying $A_t$ between 440 and
1080 GeV.
Unlike in fig. 1 we have increased the $SU(2)$ gaugino mass $M_2$ along with
\mst\, so that the single stop decays (\ref{e3a},\ref{e3b}) remain
kinematically forbidden.

\vspace*{5mm}
\item  
Cross section for \sigst\ production at the SSC. The solid line shows the
total cross section, and the dashed curve the total cross section multiplied
with $Br(\sigst \rightarrow \gamgam)$. The dotted curve is a parametrization
of the estimate of ref.\cite{15} for the minimum $\sigst \rightarrow \gamgam$
cross section detecable after 1 year of SSC running; we have extrapolated the
result of ref.\cite{15} into the region $m(\sigst) > 200$ GeV.

\end{enumerate}
\end{document}